\documentclass[fontsize=11pt,paper=a4]{scrartcl} 

\usepackage[english]{babel}
\usepackage{graphicx}
\usepackage[T1]{fontenc}
\usepackage[ansinew]{inputenc} 
\usepackage{textcomp} 
\usepackage{nicefrac}
\usepackage{natbib}
\usepackage{setspace} 
\usepackage[pdfstartview={XYZ 0 0 0.79}]{hyperref} 
\usepackage{url} 
\usepackage{booktabs} 
\usepackage{amsmath} 
\usepackage{amssymb} 
\usepackage{color} 
\usepackage{geometry} 
\usepackage{hyperref} 
\usepackage{caption} 

\title {Four-terminal magneto-transport in graphene p-n junctions
created by spatially selective doping}
\author{Timm Lohmann\footnote{t.lohmann@fkf.mpg.de} , Klaus v. Klitzing, Jurgen H. Smet}
\date{\small Max-Planck-Institut für Festkörperforschung, Heisenbergstr. 1, 70569 Stuttgart, Germany}

\begin{document}

\maketitle 
\begin{spacing}{1.5} 

\begin{abstract} In this paper we describe a graphene p-n junction
created by chemical doping. We find that chemical doping does not
reduce mobility in contrast to top-gating. The preparation technique
has been developed from systematic studies about influences on the
initial doping of freshly prepared graphene. We investigated the
removal of adsorbates by vacuum treatment, annealing and
compensation doping using NH$_{3}$. Hysteretic behavior is observed
in the electric field effect due to dipolar adsorbates like water
and ${\rm NH_3}$. Finally we demonstrate spatially selective doping
of graphene using patterned PMMA. 4-terminal transport measurements
of the p-n devices reveal edge channel mixing in the quantum hall
regime. Quantized resistances of $h/e^{2}$, $h/3e^{2}$ and
$h/15e^{2}$ can be observed as expected from theory.
\end{abstract}

Since its discovery in 2004~\cite{Novoselov1} graphene, a single
layer of hexagonally arranged carbon atoms, has developed into a
fascinating material for research. Graphene's unusual band structure
allows electric field tuning of both carrier type and concentration.
The easily accessible surface offers unique possibilities for
fabricating low dimensional systems. Chemical doping with gases has
for instance been demonstrated in 2007~\cite{Schedin}. With local
gates on top of graphene~\cite{Huard1, Oezyilmaz} transport
measurements in p-n junctions have been reported~\cite{Williams}.
Theoretical studies have predicted among others phenomena such as
Klein tunneling~\cite{Katsnelson} and Veselago lensing in ballistic
p-n junctions~\cite{Cheianov1}.

Controlling the intrinsic doping as well as the capability to change
the carrier density locally are very important issues for graphene
devices. In previous works it has been shown that graphene is
extremely sensitive to molecular adsorbates~\cite{Schedin} and top
gates have been used to control the carrier density in selected
areas of the sample~\cite{Huard1, Oezyilmaz}. The tuning capability
of a top-gated p-n junction goes at the expense of reduced sample
quality. The mobility drops due to the presence of the top-gate
oxide. Here we demonstrate a different technique based on chemical
doping to change the carrier density in selected areas of the
graphene samples. As we show later chemical doping does not reduce
the sample's mobility if executed at room-temperature. Initially, we
investigated the influence of molecular adsorbates on the field
effect behavior of graphene to gain a better understanding on how to
manipulate the adsorbates. For spatial control resist masks have
been used which can be easily patterned with e-beam lithography.
Parts of a graphene sample are covered by the resist while the
uncovered parts can be exposed to dopants like NH$_{3}$. This
technique allows us to create carrier density steps and hence
provides an alternative route to fabricate p-n junctions. The
overall density can be tuned with the help of a global back-gate.
The electronic transport properties of these p-n devices were
investigated in a four terminal geometry as a function of the
magnetic field and carrier density. In the quantum Hall regime, our
measurements yield evidence for edge channel equilibration similar
to what has been demonstrated in a two terminal geometry~\cite
{Williams}.

\begin{figure}[htbp] \centering
\includegraphics{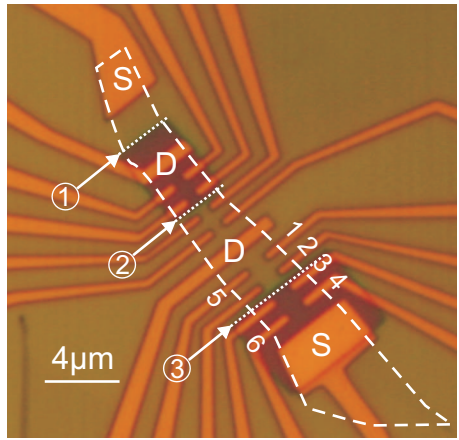} \caption{A contacted graphene
mono layer with a patterned layer of PMMA on top. The white dashed
line highlights the edges of the graphene flake. Two windows in the
PMMA allow to selectively dope the flake underneath in these
regions. This procedure results in three independent p-n junctions
(numbers in circles) for this particular design. Contacts used as
source (S), drain (D) and voltage probes (1 to 6) are labeled.}
\label{fig:windows}
\end{figure}

Individual graphene flakes were prepared on highly n-type doped
Si-substrates with a 300\,nm thick thermally grown SiO$_{2}$ layer
at the top using a micromechanical cleavage method, similar to the
one described in Ref.~\cite{Novoselov1}. Monolayers were identified
with the help of optical microscopy~\cite{Blake} and Raman
spectroscopy~\cite{Ferrari}. Cr/Au (3nm/30nm) contacts were
patterned via electron beam lithography. The global density of the
sample can be changed by applying a voltage to the backside of the
doped Si-substrate. Four-terminal resistance measurements were
performed in a Hall bar configuration with a sinusoidal source-drain
current of 100\,nA. A typical contact geometry used to explore
spatially selective chemical doping is shown in \ref{fig:windows}.
The present paragraph focuses on similarly contacted graphene
samples but without partial PMMA (Polymethylmethacrylate) coverage.
The sample was mounted in a sample stick with the following
capabilities: (i) The sample space can be evacuated down to
10$^{-5}$\,mbar. (ii) Gases such as NH$_3$, Ar, He and O$_2$ can be
introduced in a controlled fashion. (iii) The sample can be heated
up to 420\,K. In addition, it is possible after heat or gas
treatment to transfer the sample into helium atmosphere and cool it
down to 1.5\,K without any exposure to air. The resulting
experimental set-up can be used to chemically dope the flake,
control the value of the doping concentration in situ at room
temperature (RT) and subsequently freeze it in the cryostat to carry
out low temperature transport studies in magnetic fields up to
12\,T.

\begin{figure}[htbp] \centering \includegraphics{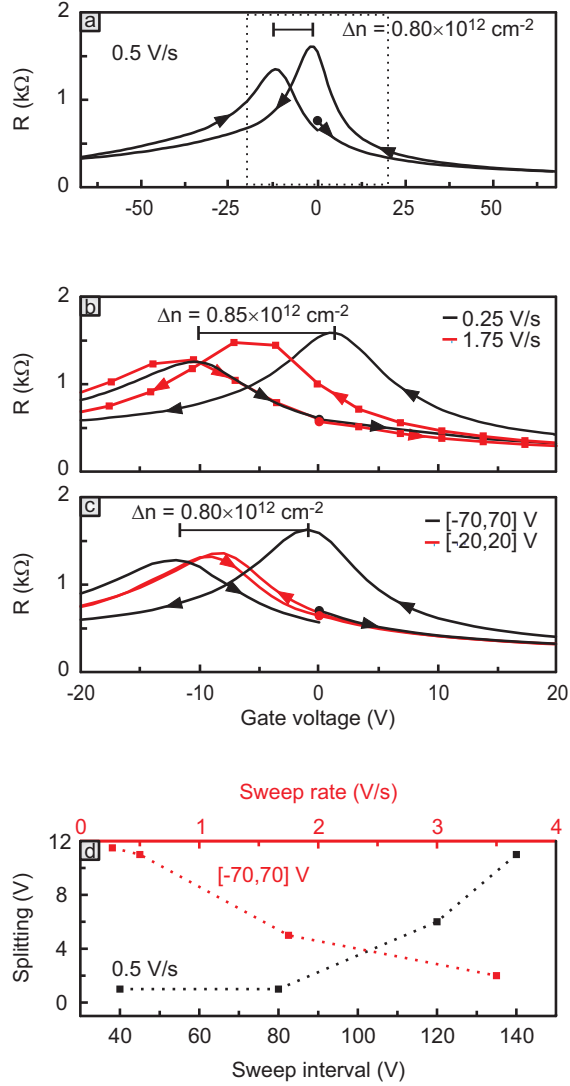}
\caption{(a) Hysteresis in the field effect behavior of a low doped
graphene monolayer under ambient conditions. The measurement was
recorded by starting from zero back-gate voltage (black-dot), sweeping
to +70 V, then -70 V and back to zero. The hysteresis is attributed
to dipolar adsorbates such as ${\rm H_{2}O}$. The arrows indicate
the sweep direction. The measurement was executed at a sweep rate of
0.5\,V/s. The dotted rectangle marks the gate voltage span plotted
in panels b and c. (b) Field effect hysteresis for two different
gate voltage sweep rates: 0.25\,V/s and 1.75\,V/s. The gate voltage
is swept in the range of [-70, 70]\,V. (c) The same measurement as
in b but here the voltage span serves as the parameter:[-20, 20]\,V
and [-70, 70]\,V. The sweep rate is equal to 0.5\,V/s. (d) Splitting
as a function of the voltage interval (bottom x-axis in black) and
sweep rate (top x-axis in red). The red/black dotted lines are
guides to the eye.} \label{fig:hysteresis}
\end{figure}

\begin{figure}[htbp]\centering \includegraphics{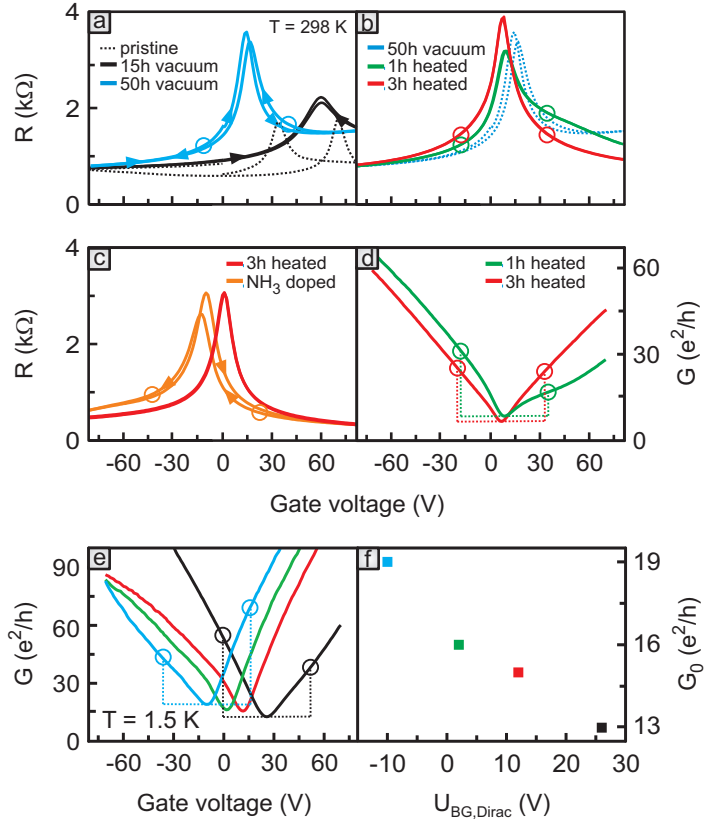}
\caption{(a) Vanishing hysteresis after 15\,hours in vacuum at
3$\cdot$10$^{-5}$ mbar (black curve) for a highly doped sample.
Reduced intrinsic doping and hysteresis after 50\,hours in vacuum
(blue curve). The black dotted line shows the initial state of the
sample after preparation. (b) Field effect behavior after 1\,hour
(green curve) and 3\,hours (red curve) annealing at 420\,K and
10$^{-5}$ mbar. The blue dotted line is the curve from (a) after
50\,hours in vacuum. (c) Shifting the charge neutral point by
chemical doping using NH$_{3}$. The initial curve (red) is measured
after 3\,h annealing. A hysteresis appears due to the dipolar nature
of NH$_{3}$ (similar to that caused by water but with opposite
direction). (d) Conductance plot of the green and red curve from
\ref{fig:conditions}b to bring out better the difference between
electron and hole conduction. All measurements in (a) to (d) were
performed at $\approx$298\,K. (e) Influence of doping on the e-h
asymmetry. The black curve shows the conductance of a freshly
prepared sample at 1.5 K. The red, green and blue curve are measured
successively after an extra dose of 0.1\,mmol NH$_{3}$ was
introduced into the sample chamber at room temperature. One hour
after each additional gas exposure, the sample was cooled back down
to 1.5\,K for the measurement in order to freeze out the hysteresis.
The equidistant shift of $\approx12\,V$ to the left between the
curves confirms that ${\rm NH_3}$ compensates the initial p-doping.
Note that the asymmetry between electron and hole conduction is
inverted upon doping with ${\rm NH_3}$ (compare the black and blue
circles). (f) Plot of the conductance G$_{0}$ at the charge neutral
point taken from the data in panel e versus U$_{BG, Dirac}$.}
\label{fig:conditions}
\end{figure}

Our studies first address the room temperature field effect
characteristics on freshly prepared flakes under ambient conditions.
The four terminal sample resistance exhibits strong hysteretic
behavior. When comparing up and down sweeps of the back-gate voltage
the resistance reaches a maximum at different back-gate voltages for
the two sweep directions as illustrated in \ref{fig:hysteresis}a.
The peaks are shifted by as much as 10\,V which corresponds to a
density difference of 0.8$\cdot$10$^{12}$\,cm$^{-2}$. For samples
with higher initial doping the hysteresis or splitting is even
larger (for instance \ref{fig:conditions}a, dotted curve). The
observed behavior is known from carbon nanotubes~\cite{Kim} and was
attributed to dipolar adsorbates such as water which act as charge
traps. When executing subsequent up and down sweeps, small
deviations from the first cycle are observed. The splitting however
stays constant. This has been verified for up to 10 cycles. Since
charge trapping by water is a dynamic process it depends on the
sweep conditions of the applied gate voltage. Low sweep rates result
in a large splitting of the resistance peaks as shown in
\ref{fig:hysteresis}b. We display curves for two different values of
the sweep rate: 0.25\,V/s and 1.75\,V/s. A second relevant parameter
is the sweep interval, i.e. the amount of charge injected into the
sample. A large voltage span such as [-70V, 70V] produces a large
splitting. \ref{fig:hysteresis}c displays data recorded for two
gate-voltage sweeps: from -70 to 70\,V and -20 to 20\,V. Other sweep
conditions yield a consistent dependence as shown in
\ref{fig:hysteresis}d.

In order to influence the sample's initial doping and the hysteresis
described above we treated samples under various conditions. The
three important parameters, which were explored, are (i) vacuum down
to 10$^{-5}$\,mbar, (ii) temperatures of about 420\,K and (iii) gas
atmosphere in the sample chamber. Changes in the sample quality and
doping by varying these parameters were assessed by investigating
the following sample characteristics: (1) The existence of
hysteresis in the field effect behavior, which can be caused by the presence
of dipolar species. (2) The back-gate voltage at which charge
neutrality is achieved. It serves as an indicator for the amount of
(unintentional) doping. (3) The resistance/conductance at the charge
neutrality point. It reflects the degree of density inhomogeneity in
the sample. (4) Finally, the mobility difference between Dirac
electrons and holes. It is governed by the type and number of doping
adsorbates. In the following paragraphs, we will discuss how each of
these characteristics change.

{\em Vacuum treatment:} Keeping a freshly prepared sample under
3$\cdot$10$^{-5}$\,mbar at room temperature for several hours
strongly changes the field effect behavior. It diminishes the
hysteresis and frequently also causes a shift of the charge
neutrality point closer to zero back-gate voltage. An example is
shown in \ref{fig:conditions}a. The hysteresis has nearly vanished
after pumping down to 3$\cdot$10$^{-5}$\,mbar for 15 hours (black
curve). The charge neutrality point lies in between the peaks of the
up and down sweep recorded prior to sample evacuation (black dotted
line for comparison). After continued pumping for a total of
50\,hours (blue curve) a significant shift of the charge neutrality
point is observed. Presumably, this shift is caused by the gradual
desorption of doping adsorbates in vacuum. Since no elevated
temperatures are required to induce a change, it indicates that at
least one adsorbed species that causes hysteresis and p-doping is
only loosely bond to the graphene surface. The desorption is also
accompanied by an increase of the resistance at the charge
neutrality point from $\approx2\,k\Omega$ to $\approx3\,k\Omega$.
The non-zero conductivity at charge neutrality likely originates
from the existence of electron and hole puddles as demonstrated in
scanning SET (single electron transistor) experiments~\cite{Martin}.
Even though on average the sample is charge neutral, locally charge
carriers are available and Klein tunneling ensures that transport
across puddles of opposite charge polarity is not impeded. The
increase of the resistance at the charge neutrality point suggests
that fewer charge carriers are available in the electron and hole
puddles because of a reduced amplitude of the density fluctuations.
Finally, note that the blue curve after 50\,hours of vacuum
treatment exhibits a pronounced asymmetry between the electron and
hole conduction (blue circles mark two voltages where the electron
and hole density are equal). These observations will be discussed in
more detail below.

{\em Sample annealing:} Heat treatment is the second possibility to
remove dopants. It is more effective than pumping and requires less
time. Heat treatment improves the sample quality in terms of our
four criteria: smaller hysteresis, a charge neutrality point closer
to zero back-gate voltage, a larger resistance at the charge
neutrality point and a smaller mobility difference between electrons
and holes. This is in agreement with previous reports (see. Moser et
al.~\cite{Moser1}, Ishigami et al.~\cite{Ishigami}). We note that
heating should proceed under vacuum conditions, but a long term
initial pumping is not needed to reach the end result. Pumping alone
might be useful for devices which should not be exposed to elevated
temperatures. Here we illustrate how the sample that underwent the
50\,hour evacuation reacted to the heat treatment
(\ref{fig:conditions}b). The hysteresis has vanished entirely after
the anneal. The charge neutrality point has not shifted much
further. The difference between the electron and hole mobility
remains present after a 1\,hour anneal (green circles), but has
disappeared after 3\,hours of annealing (the red circles have no
vertical offset). The resistance at the charge neutrality point
slightly dropped after 1\,hour of heat treatment, but increased to
about $4\,k\Omega$ after a 3\,h anneal at 420\,K in vacuum. These
observations all suggest that most of the initially adsorbed species
are removed after annealing. (i) When dipolar adsorbates are
effectively removed during the anneal, the hysteresis will vanish.
(ii) When p-dopants such as water and oxygen disappear during the
anneal, the charge neutrality point will shift close to zero
back-gate voltage. (iii) The amplitude of the density fluctuations
in the sample may drop upon removal of doping adsorbates. It will
increase the resistance at the charge neutrality point.

{\em Gas exposure:} The amount of doping can of course also be
manipulated by exposing the sample to suitable gases. In
\ref{fig:conditions}c we illustrate doping with NH$_{3}$. The red
curve has been measured on a sample similar to the one described
previously after a 3\,hour anneal. NH$_{3}$ acts as an n-type dopant
and indeed the charge neutrality point shifts to negative back-gate
voltages (orange curve) after the flake has been exposed. Hence,
part of the initial p-doping is compensated. Note that a small
hysteresis reappears with opposite sign as in \ref{fig:hysteresis}.
It is attributed to the dipolar nature of the NH$_{3}$ molecules. An
asymmetry between the electron and hole side is also clearly visible
in the orange curve (orange dots), but it is distinct from the
asymmetry observed on flakes that were not treated with ${\rm NH_3}$
in \ref{fig:conditions}a and b. Untreated flakes exhibit superior
hole conduction, while the ${\rm NH_3}$ treated case shows better
electron than hole conduction. \ref{fig:conditions}e illustrates the
evolution of the field effect behavior when gradually exposing a
flake to a larger amount of ${\rm NH_3}$. The black solid line
represents the conductance obtained on the as prepared flake at
1.5\,K. The sample was then warmed up in vacuum to room temperature
and exposed to a dose of 0.1\,mmol of ${\rm NH_3}$. After cooling
back down to 1.5\,K, the field effect behavior was recorded again.
The same procedure was repeated several times. For each additional
gas exposure the charge neutrality point shifted by approximately
-12\,V confirming that ${\rm NH_3}$ molecules act as n-type dopants.
The conductance at the charge neutrality point increases with
increasing gas exposure. It is plotted in panel f as a function of
the back-gate voltage at which charge neutrality is reached. These
data suggest that spatial density fluctuations are enhanced upon
doping, which would imply that the ${\rm NH_3}$ molecules get
preferentially adsorbed in certain regions only and not
homogeneously across the entire flake. This issue may be intimately
connected with the memory effect observed when flakes, which were
treated in vacuum are exposed back to air. As we showed above, the
vacuum and heat treatment drastically changes the four
characteristics. They however return to their initial values (within
a tolerance of a few $\%$ only) after a short time ($< 1\ {\rm
min}$) when exposing the flake back to air irrespective of the
duration and sequence of treatments. This reversibility adds support
to the argument that doping adsorbates preferentially attach to
surface and edge specific locations of the pristine flake. A similar
argument was invoked previously in Ref.~\cite{Moser2}. A theoretical
description of how dopants may affect the conductance at the charge
neutrality point has been reported by Cheianov et
al.~\cite{Cheianov2} and is based on percolation in a random network
of resistors.

The experiment with gas exposure in~\ref{fig:conditions}e also
convincingly demonstrates the polarity dependent impact of doping
adsorbates on the mobility of charge carriers. For the as prepared
flake with substantial initial p-doping due to water and oxygen for
instance, hole conduction excels over electron conduction when
comparing the conductance values for identical electron and hole
densities. The same was already seen in panels a-d (in particular
panel d, where the resistance data of panel b is replotted on a
conductance axis). However, upon exposing the flake to more and more
${\rm NH_3}$, electron conduction becomes superior. The conductance
at identical carrier densities is highlighted for one particular
density with circles for some curves in the different panels to ease
the comparison. The development of this asymmetry between the
electron and hole side is in qualitative agreement with recent
theoretical work~\cite{Robinson}. It was attributed to the energy
dependent scattering potential generated by molecular adsorbates.
The scattering potential suppresses conductance on one side of the
Dirac point depending on the n- or p-type character of the doping
adsorbate, while for the other side conductance is only weakly
influenced. The deposition of metallic contacts has also been held
responsible for the appearance of an electron-hole asymmetry in
Ref.~\cite{Huard2}. These authors have invoked charge-density
pinning underneath contacts to explain this asymmetry. Scanning
photocurrent microscopy experiments~\cite{Lee} have indeed verified
that deposited metals cause a doping of graphene and a density
pinning in the vicinity of contacts. This doping in turn may be a
cause of the observed asymmetry in Ref.~\cite{Huard2}. The sign of
the asymmetry observed in our experiments for p- and n-doping is
consistent with the I-V measurements for p- (Au) and n-doping (Ti)
metals in Lee et al.~\cite{Lee}. Note that the removal of adsorbates
upon vacuum and heat treatment largely removes the mobility
difference between electrons and holes (\ref{fig:conditions}b and c,
red curves for instance) as anticipated for ideal graphene in view
of the intimate connection between Dirac electrons and holes.
Although the asymmetry can be removed we do not observe a
significantly increasing mobility after annealing. The mobilities of
our freshly prepared samples lie between 5000 and 8000\,cm$^{2}$/Vs.
Under the PMMA we have mobilities between 3000 and
5000\,cm$^{2}$/Vs.

In the following section we demonstrate that chemical doping can be
exploited to study graphene p-n junctions when masking part of the
graphene layer with a resist. In a first step, ohmic contacts are
made to a graphene monolayer using electron beam lithography.
Subsequently, the sample is coated with a PMMA resist layer of
200\,nm thickness. A second electron beam lithography step opens
windows in this resist layer. The sample is then mounted in the
sample holder with vacuum chamber. Exposure to any gas, which dopes
either $n$- or $p$-type, creates a density step at each boundary
between the exposed graphene and those regions of the graphene flake
still covered by PMMA. \ref{fig:windows}, discussed briefly
previously, shows an optical image of such a flake with two windows
in the PMMA layer on top. The flake itself is demarcated by the
white dashed line. This device contains three such p-n junctions
(dotted lines), which are labeled with the encircled numbers.
Junction 1 makes up a two terminal geometry, while the others can be
measured in a four terminal configuration. All data shown in this
section have been acquired using the voltage probes 1 through 6 of
junction 3. The two other devices revealed similar behavior.

\begin{figure}[htbp] \centering \includegraphics{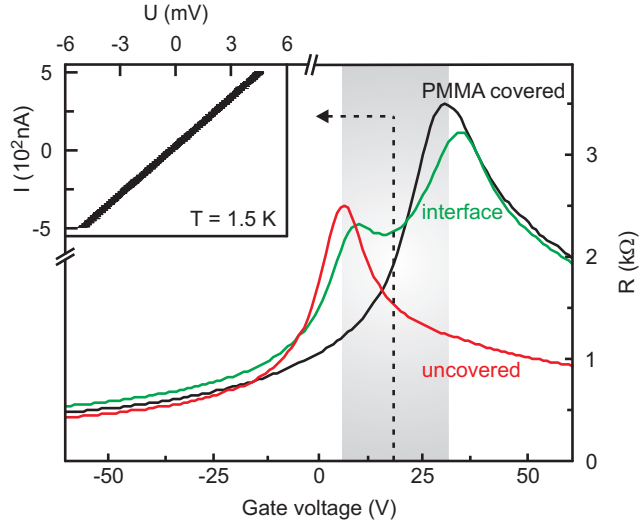}
\caption{Field-effect measurement of a device as shown in
\ref{fig:windows}. The black and red curves display the field-effect
behavior in the PMMA-covered  and  uncovered  regions. The two
regions exhibit a density difference of about $1.8\cdot10^{12}\ {\rm
cm}^{-2}$. The green curve was recorded across the interface of the
two regions and is a superposition of the black and red curves. By
choosing a back gate voltage in between the charge neutrality points
of both regions a p-n junction is formed. The p-n region has been
highlighted with a gray shading. The inset shows the I-V
characteristics of such a p-n interface for a backgate voltage of
19\,$V$. As theoretically predicted \cite {Katsnelson} the device
shows no "diode-like" behavior but rather Ohm's law.}
\label{fig:gradient}
\end{figure}

After mounting the sample in the sample holder with vacuum chamber,
it was annealed in vacuum at 420\,K in order to remove residual
adsorbates on those parts of the flake not covered by resist. The
sample chamber evacuation and the heat treatment already introduce a
doping step across the flake, since adsorbates underneath the PMMA
can either not be removed or diffuse only slowly. The carrier
density of the covered and uncovered areas can be monitored in situ
by measuring the resistance in both regions as a function of the
back-gate voltage during pumping and annealing in vacuum at room
temperature. After this procedure the sample is transferred into the
cryostat without exposure to air. \ref{fig:gradient} displays such
field effect measurements for the PMMA covered (black line, voltage
probes 1 and 2) and the exposed part of the graphene flake (red
line, voltage probes 3 and 4) after evacuation and annealing. The
charge neutrality points are reached at back-gate voltages of
$V_{\rm Dirac}^{\rm covered}= 32\,V$ and $V_{\rm Dirac}^{uncovered}=
9\,V$. For the sake of completeness we also show the longitudinal
resistance measured across the junction (green line, voltage probes
2 and 3). The difference in the charge neutrality points corresponds
to a density difference of $1.8\cdot10^{12}\ {\rm cm}^{-2}$. By
choosing a back-gate voltage in between (19\,$V$) a p-n junction is
created.
The inset in \ref{fig:gradient} depicts the $I-V$ characteristic
measured between the source and drain contact across the p-n
interface, which forms at a back-gate voltage $V_{\rm bg} = 19\,V$.
The measurements were carried out at 1.5\,K. The $I-V$ trace does
not resemble that of a conventional p-n diode in accordance with
theory~\cite{Katsnelson}. In a conventional semiconductor charge
carriers have to overcome a bandgap in order to reach the conduction
band. Therefore a conventional p-n junction is a diode in contrast
to that in gapless graphene. But the missing gap is not the only
reason for the absence of a diode-like I-V curve in graphene. A
deeper reason is the conservation of the pseudospin which acts as an
additional quantum number. It is a consequence of graphene's unusual
bandstructure. Pseudospin conservation allows charge carriers in
graphene to tunnel through a potential barrier like a p-n interface
with 100\% efficiency at normal incidence, independent of the width
and height of the barrier. A gapless material with a parabolic
bandstructure would show an oscillating transmission coefficient
between zero and one while that of graphene is always one.

Therefore the measured $I-V$ characteristic is ohmic with a
resistance of 9\,$k\Omega$. The device geometry has approximately
1.5 squares. In view of the two terminal configuration, this
resistance still contains contributions from the metal
contact-graphene interfaces. From a comparison with
four-terminal resistance measurements, we estimated the contact
resistance to be below 1\,k$\Omega$.

\begin{figure}[htbp] \centering \includegraphics{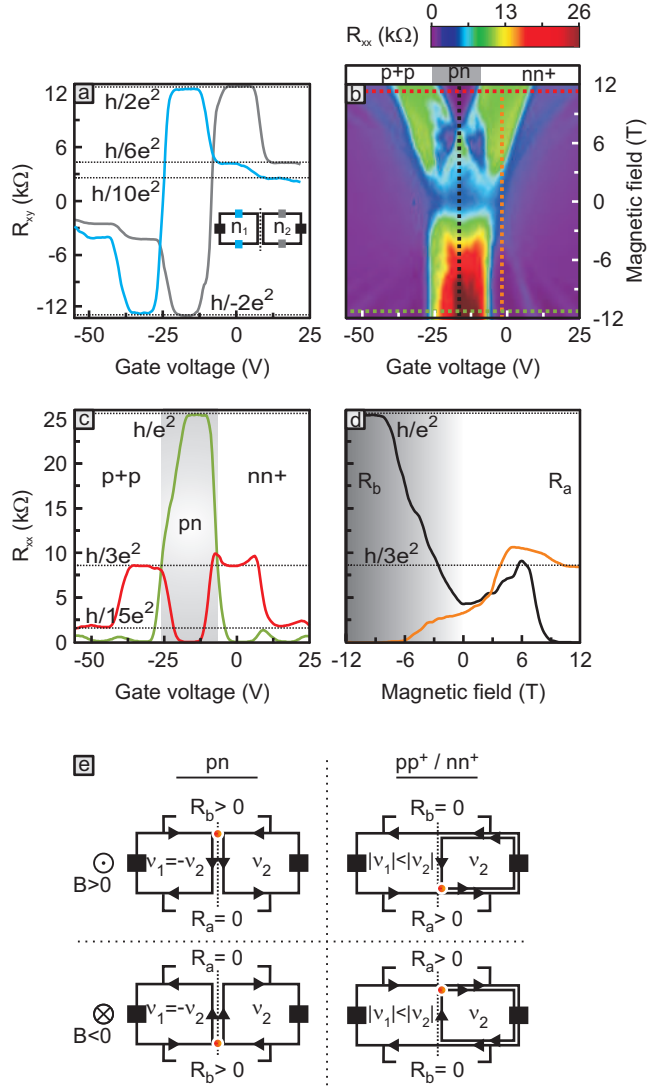}
\caption{Low temperature (1.5\,K) magneto transport behavior of a
graphene p-n junction created by selective chemical doping. (a) Hall
resistance $R_{xy}$ in two regions with different doping measured as
a function of the gate voltage. The magnetic field is 12\,T. The
doping difference of about 15\,$V$ ($1.2\cdot10^{12}\,cm^{-2}$)
provides a precise overlap of neighboring filling factors. The
drawing illustrates the device geometry. (b) Color rendition of the
longitudinal resistance $R_{xx}$ as a function of the magnetic field
and the gate voltage. The dotted lines belong to
\ref{fig:magneto}c,d. (c) Plot of the horizontal line sections
from \ref{fig:magneto}b. The characteristic resistance quanta
$h/e^{2}$, $h/3e^{2}$ and $h/15e^{2}$ are indicated by black dotted
lines. (d) Plot of the vertical line sections from
\ref{fig:magneto}b. The grey/white region highlights the regime
before/after edge channel interaction. (e) Schematic of edge channel
mixing in 4-terminal geometry for the bipolar (p-n) and homopolar
(p+p/nn+) case and for positive and negative magnetic field. The
orange dots show where edge channel equilibration most likely takes
place in the two cases.} \label{fig:magneto}
\end{figure}

The graphene p-n junctions fabricated by spatially selective
chemical doping were investigated further in the quantum Hall
regime. In graphene one expects an unconventional integer quantum
Hall effect~\cite{Novoselov2, Zhang1} due to the presence of a
Landau level at zero energy (Dirac point), which is shared by both
electrons and holes. Each Landau level is four fold degenerate due
to the valley and spin degree of freedom. Therefore, when the spin
and valley degeneracies are not lifted the system condenses in
integer quantum Hall states when the filling factor $\nu$, i.e.~the
number of filled Landau levels, equals $2, 6, 10,\ldots$ and $-2,
-6, -10,\ldots$ with Hall resistance $R_{\rm xy} = h/(e^2 \nu)$.
Negative filling factors correspond to hole fillings. In order to
avoid parasitic effects from folded/misshaped edges we use graphene
devices patterned into a Hall bar shape with e-beam lithography and
oxygen plasma etching for these experiments. Furthermore the Hall
bar geometry has the advantage that the metal/graphene interfaces of
the contacts can be far away from the measured region. The reported
doping due to metal deposition~\cite{Lee} has then no influence for
a measurement of the longitudinal resistance itself. The PMMA
masking and contact scheme is similar to the device shown in
\ref{fig:windows}. \ref{fig:magneto}a shows the Hall resistance
$R_{xy}$ of two separate regions with different densities $n_{1}$
and $n_{2}$. The shift between the charge neutrality points, i.e.
the resistance maxima, of about 15\,$V$ corresponds to a density
difference $\Delta n = |n_2 - n_1| = 1.2\cdot10^{12}\ {\rm
cm}^{-2}$. This value is not arbitrary but was chosen carefully in
order to ensure that the two regions with different density
simultaneously condense in distinct integer quantum Hall states. The
following combinations of integer filling factors $(\nu_1,\nu_2)$
can be reached (see \ref{fig:magneto}a) as the back-gate voltage is
swept: (10,6), (6,2), (2,-2), (-2,-6) and (-6,-10). Hence p$^+$p,
nn$^+$ as well as p-n junctions can be formed.

When the p- and n-regions condense in a quantum Hall state, the bulk
turns incompressible and chiral edge channels flow along the
boundary in opposite directions for the p and n-regions. Previous
experimental work on two-terminal p-n junctions in a strong
perpendicular magnetic field has shown that under quantum Hall
conditions for both the p and n-region edge channels equilibrate
efficiently when they meet at the p-n interface~\cite{Williams}. It
gives rise to two terminal resistances, which are properly predicted
within the Landauer-Büttiker edge channel picture under the
assumption of full equilibration~\cite{Abanin}. While these previous
studies focused on two terminal devices, here we address this
physics in the four terminal configuration. Apart from removing the
contact resistance contribution, a four terminal arrangement offers
the advantage that one can measure voltage drops across the junction
before and after edge channels interacted on the opposite boundaries
of the sample as schematically depicted in \ref{fig:magneto}e. For
the p-n regime, the Landauer-Büttiker formalism predicts
\begin{eqnarray}\label{eq:mixing1}
R_{b}=\frac{h}{e^{2}}\left(\frac{1}{|\nu_{1}|}+\frac{1}{|\nu_{2}|}\right)\
\ \ \ {\rm and} \ \ R_{a} = 0
 \end{eqnarray}
\textbf{before} ($R_{b}$) and \textbf{after} ($R_{a}$) edge channels
interact along the junction. If the carrier type is the same on
either side of the junction, these resistances read
\begin{eqnarray}\label{eq:mixing2}
   R_{b} = 0\ \ \ \ {\rm and} \ \ R_{a} = \frac{h}{e^{2}}\left(\frac{1}{|\nu_{1}|}-\frac{1}{|\nu_{2}|}\right)
 \end{eqnarray}
for the p$^+$p or nn$^+$-regime. $\nu_{1}$ and $\nu_{2}$ are the
filling factors in the adjacent regions of the sample. If in both
regions the charge carriers possess the same polarity, edge channels
do not interact along the junction itself, but rather when those
originating from the region with a lower absolute filling factor
have crossed the junction (right panel of \ref{fig:magneto}e). The
above description assumes full equilibration of the electrochemical
potentials of the edge channels where they interact. As illustrated
in the lower row of panels in \ref{fig:magneto}e, inverting the
direction of the magnetic field interchanges the Hall bar side where
the interaction of edge channels did not take place yet with the
Hall bar side where edge channels have interacted. The formulas for
$R_{b}$ and $R_{a}$ remain unaffected. Note that measuring the
longitudinal resistance on opposite sides of the Hall bar is
equivalent to inverting the sign of the magnetic field.

A color rendition of the longitudinal resistance $R_{xx}$ as a
function of the magnetic field and the back-gate voltage is shown in
\ref{fig:magneto}b. These data allow us to verify the above picture.
A hallmark of a four terminal Quantum Hall effect (QHE) experiment
on a sample with regions of different filling factors is the
magnetic field-asymmetry of $R_{\rm xx}$. In a graphene device
without a junction, the measurement of $R_{\rm xx}$ would yield a
magnetic field-symmetric Shubnikov-de Haas fan since there is no
distinction between the resistances measured on opposite sides of
the sample.
In the p-n junction here, which forms for back-gate voltages between
-5 and -25\,V, one observes zero resistance for fields between 8 to
12\,T, while a resistance of approximately 25.5\,k$\Omega$ is
recorded at negative fields -8 to -12 T (or equivalently on the
opposite edge of the sample). These two cases correspond to the top
and bottom left panels in \ref{fig:magneto}e when measuring the
resistance across the junction using voltage probes at the bottom of
the Hall bar geometry. In the top left panel for positive values of
the magnetic field, the edge channels originating from the source
and drain contact have equilibrated their electrochemical potentials
as they are jointly running along the junction. There is no longer a
difference in the electrochemical potentials when measuring at the
bottom. The recorded resistance is zero. For negative values of the
magnetic field, edge channel potentials are probed prior to
equilibration along the p-n interface. Hence, the resistance is
determined by the potential difference applied across the source and
drain contacts. Since $\nu_1 = -2$ and $\nu_2 = 2$ we anticipate
$R_{b} = h/e^{2}$ (Eq.~1) or 25812.807\,$\Omega$. Our measured value
of $25.5\ {\rm k\Omega}$  is within the precision of our
experimental setup.

When the system condenses either in a combination of the (-6,-2) or
(2,6) integer quantum Hall states, the resistance drops to zero in
both cases. This is again in line with the above edge channel
picture. As seen in the bottom right panel of \ref{fig:magneto}e,
even though a different number of edge channels enters each of the
voltage probes, these edge channels originate all from the same
contact. Hence, the voltage probes float to the same electrochemical
potential. For positive values of the magnetic field (top right
panel), the region with filling factor $\nu_1$ ($|\nu_1| < |\nu_2|$)
can support fewer edge channels. The extra edge channels which exist
in the region with filling factor $\nu_2$ get reflected at the
boundary (top right panel of \ref{fig:magneto}e). After running
along the junction they mix with channels originating from the other
contact. From equation \ref{eq:mixing2} we expect
$h/3e^{2}$=\,8604.269\,$\Omega$ which is in agreement with the
observed value of 8535\,$\Omega$. Also for the other combinations of
integer quantum Hall states such as for instance (6,10) the edge
channel picture with full equilibration performs very well. This is
best seen in the horizontal and vertical line sections through the
data of \ref{fig:magneto}b plotted in panels c and d. Here, we find
zero resistance for the negative sign of the field and $1.67\ {\rm k
\Omega}$ or close to $h/15e^{2}$=\,1720.8538\,$\Omega$ for the
positive sign.

In summary, we have investigated the influence of vacuum, elevated
temperatures and some gases on the electronic properties of
graphene. A hysteresis in the field effect behavior of graphene
similar to what has been previously reported on carbon nanotubes,
was observed under ambient conditions. It was attributed to adsorbed
dipolar molecules such as water. Pristine graphene samples are
typically p-type doped at a level of 2$\cdot$10$^{12}$\,cm$^{-2}$.
Vacuum treatment reduces the hysteresis and also partially removes
dopants. Additional heat treatment under vacuum is more effective in
removing the doping adsorbates. It also improves the sample
homogeneity as attested by an increased value of the resistance at
the charge neutrality point, which indicates an amplitude reduction
of the electron-hole puddle landscape. Ammonia has served as an
n-type dopant. It has been used to compensate the initial p-type
doping without annealing. This chemical doping however leads to an
enhanced inhomogeneity as reflected by a reduced resistance at the
charge neutral point. An asymmetry between electron and hole
mobilities can be observed. Finally, four terminal magneto-transport
studies were performed on graphene p-n junctions fabricated by
spatially selective chemical doping. These studies revealed a strong
equilibration of edge channels near the junction.


\end{spacing}
\end{document}